\documentclass{article}
\usepackage{arxiv}

\usepackage{times}
\usepackage{url}
\usepackage[utf8]{inputenc}
\usepackage{graphicx}
\usepackage{amsmath}
\usepackage{amsthm}
\usepackage{booktabs}
\usepackage{algorithmic}
\usepackage[switch]{lineno}
\usepackage{array}
\usepackage{soul}
\usepackage{caption}
\usepackage{adjustbox}
\usepackage{amssymb}
\usepackage{algorithm}
\usepackage[
    backend=biber,
    style=numeric,
    sorting=ynt
]{biblatex}
\usepackage[hidelinks]{hyperref}

\title{FantasyID: Face Knowledge Enhanced ID-Preserving Video Generation}

\author{
  Yunpeng Zhang$^{1,2}$\footnote{Equal Contribution}\and
  Qiang Wang$^1$\footnotemark[1]\and
  Fan Jiang$^1$\footnote{Project Leaders}\and
  Yaqi Fan$^2$\and
  Mu Xu$^1$\footnotemark[2]\and
  Yonggang Qi$^2$\footnote{Corresponding authors}
  \affiliations
  $^1$AMAP, Alibaba Group\\
  $^2$Beijing University of Posts and Telecommunications\\
  \emails
  \{yijing.wq, frank.jf, xumu.xm\}@alibaba-inc.com,
  \{bryan2233, yqfan, qiyg\}@bupt.edu.cn
}

\let\oldtwocolumn\twocolumn
\renewcommand\twocolumn[1][]{
    \oldtwocolumn[{#1}{
    \begin{center}
           \includegraphics[width=17cm]{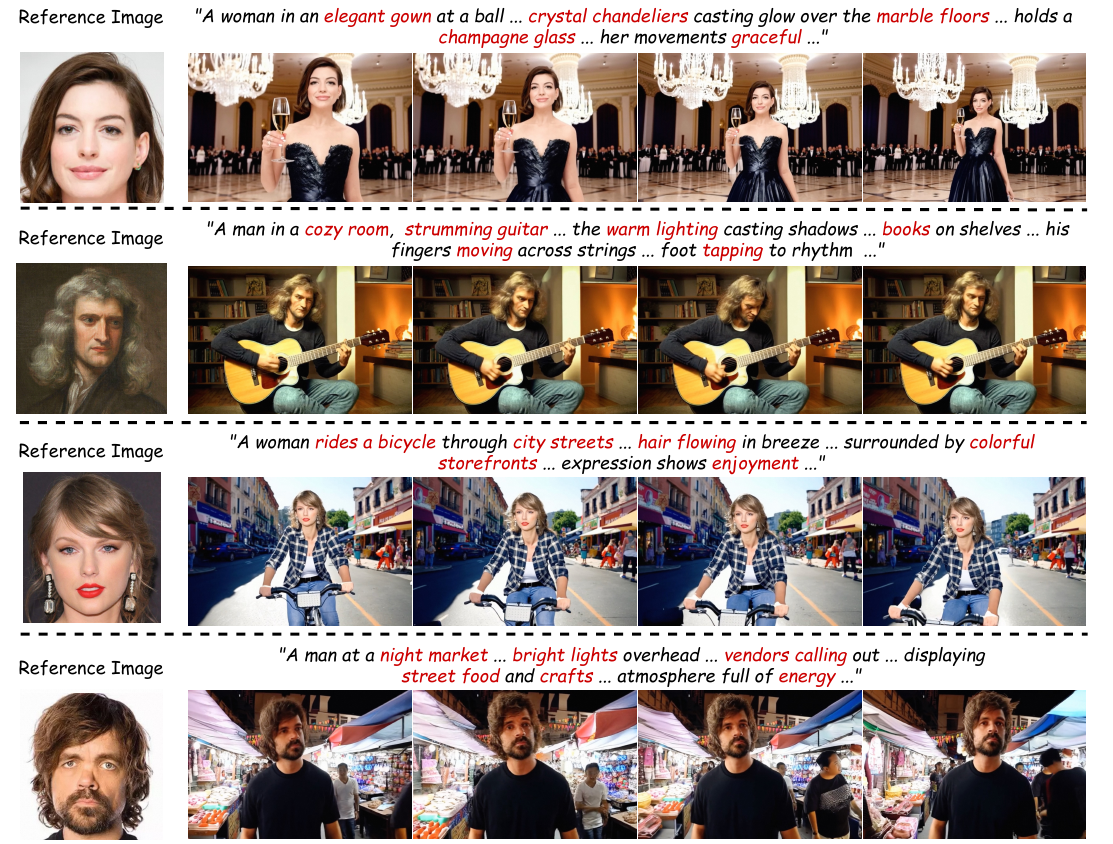}
            \captionsetup{width=16.3cm} 
           \captionof{figure}{\textbf{Examples of FantasyID.} Given a human face image, FantasyID generates ID-preserving videos with enhanced motion dynamics and more stable facial structures.}
            \label{fig:show_result}
        \end{center}
    }]
}

\begin{document}

\maketitle

\begin{abstract}
Tuning-free approaches adapting large-scale pre-trained video diffusion models for identity-preserving text-to-video generation (IPT2V) have gained popularity recently due to their efficacy and scalability. However, significant challenges remain to achieve satisfied facial dynamics while keeping the identity unchanged. In this work, we present a novel tuning-free IPT2V framework by enhancing face knowledge of the pre-trained video model built on diffusion transformers (DiT), dubbed FantasyID. Essentially, 3D facial geometry prior is incorporated to ensure plausible facial structures during video synthesis. To prevent the model from learning ``copy-paste'' shortcuts that simply replicate reference face across frames, a multi-view face augmentation strategy is devised to capture diverse 2D facial appearance features, hence increasing the dynamics over the facial expressions and head poses. Additionally, after blending the 2D and 3D features as guidance, instead of naively employing adapter to inject guidance cues into DiT layers, a learnable layer-aware adaptive mechanism is employed to selectively inject the fused features into each individual DiT layers, facilitating balanced modeling of identity preservation and motion dynamics. Experimental results validate our model’s superiority over the current tuning-free IPT2V methods. Our project page: \url{https://fantasy-amap.github.io/fantasy-id/}.

\end{abstract}

\begin{figure*}[h]
    \centering
    \includegraphics[width=18cm]{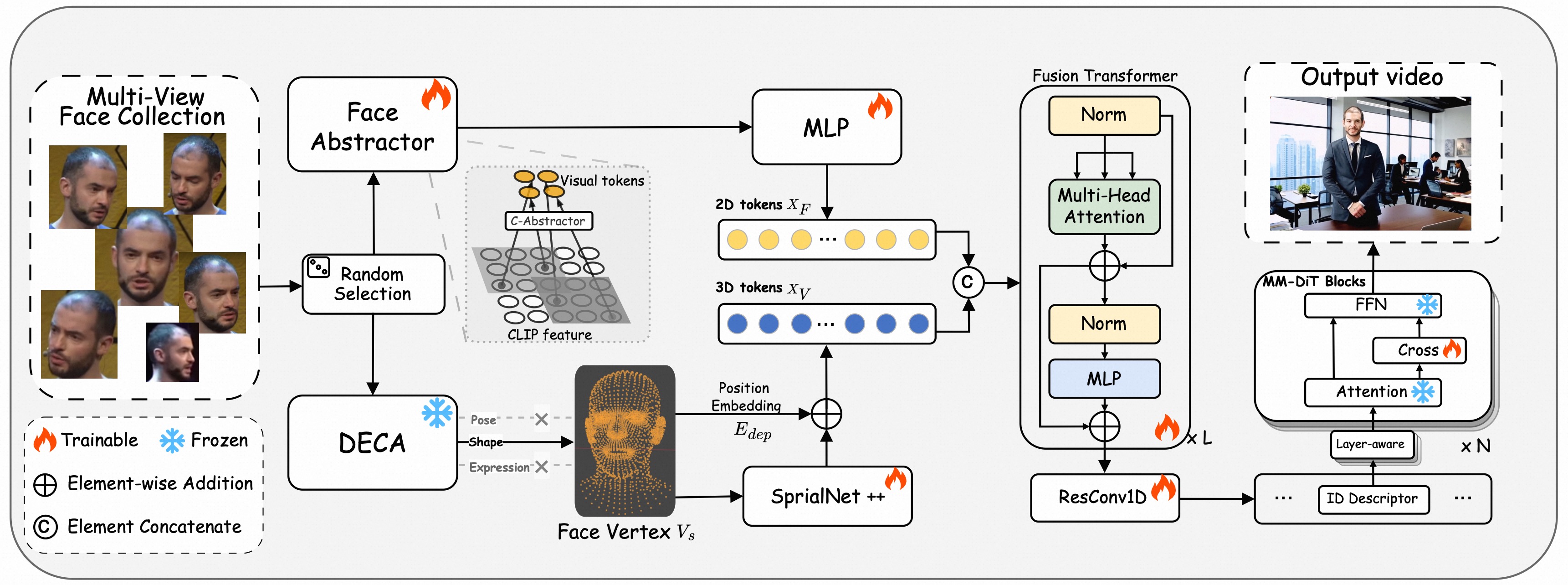}
    \caption{\textbf{Overview of FantasyID.} The framework constructs a multi-view face collection, randomly selects one face as the reference input, and employs face abstractor to extract 2D visual tokens while using DECA to extract 3D face vertex tokens. We fuse both the 2D and 3D tokens with fusion transformer layers and guide DiT-based model via a layer-aware signal injection method.}
    \label{fig:overview}
\end{figure*}

\section{Introduction}

Identity-preserving text-to-video generation (IPT2V) aims to generate videos from a reference image while consistently maintaining the identity across frames \cite{wu2024videomaker, comas2024magicmirror, ma2024magic, polyak2024movie}. Solving this problem provides valuable insights into developing compelling applications, such as personalized avatars, immersive try-ons, interactive storytelling, and more. Upon the powerful generative capability of large-scale pre-trained video diffusion models, recent works on IPT2V, such as ID-Animator \cite{he2024id} and ConsisID \cite{yuan2024identity}, shifted to model adaptation, thus avoiding case-by-case tuning during inference. 

Despite notable advancements, current identity-preserving video diffusion models face three critical challenges rooted in their architectural and training paradigms. First, current methods exhibit limited knowledge of facial structures, making them vulnerable when confronted with intricate facial movements. Besides, tuning pre-trained T2V models with a single-view reference face may encounter the "copy-paste" issue \cite{polyak2024movie, he2024id}, that excessive reliance on a monocular static image could restrict the desired diversity of facial expressions in the video. 

Moreover, the intrinsic hierarchical nature of DiT causes layer-specific sensitivity to control signals \cite{yu2024representation, yuan2024identity}, calling for dedicated conditioning strategies to harmonize identity preservation and temporal coherence throughout generation.

To tackle the first challenge, we propose integrating 3D facial geometry priors into our model to ensure stable and consistent structures of the human face during generation.

Specifically, DECA \cite{feng2021learning} is employed to extract the essential identity-specific 3D feature, i.e., a shape point cloud, which is found effective for identity preservation. The identity-irrelevant features, e.g., pose and expression, are discarded. Moreover, this introduced 3D shape prior is conveniently manageable by varying the 3D points’ locations so the generated human face could change accordingly.

To mitigate the issue of static motion, we devise a multi-view face adaptation strategy to avoid learning shortcuts that directly replicate the static face across frames in the generated video. Namely, we augment a monocular reference face image with its variants from different viewpoints, forming a face pool for the same identity obtained from the training human video. We then randomly select any of them as input for pre-trained video model adaptation.

It turns out that this can enforce the adapted IPT2V model to capture detailed diverse 2D facial appearance features, thereby improving the dynamics performance of the generated video. 

Given the obtained 3D and 2D facial features from the reference face image, we further blend them together using a transformer-based feature fusion module to guide the pre-trained video diffusion model to produce identity-specific human video. However, we figure that it is inefficient to inject the fused feature into the pre-trained DiT layers naively by adapter as it is known that the DiT lower layers tend to capture the overall structure and upper layers for details. Therefore, we introduce a layer-aware injection mechanism to allow the model to adaptively select the most beneficial cues from the fused features. 

To this end, our contributions are three-fold. (i) To our best knowledge, this is the first attempt that 3D facial priors extracted from a single-view reference image are employed to enhance the facial structure stability, thus benefiting ID preservation throughout video generation.

(ii) By employing a multi-view facial augmentation strategy, we can significantly enhance the perception of 2D facial appearance across a wide range of viewpoints, thus benefiting the motion dynamics associated with facial expression and head pose.

(iii) A learnable layer-aware feature guidance mechanism is devised to facilitate precise control for a balanced ID-preserving and dynamics modeling, offering high-fidelity human video with better temporal coherence and identity consistency. 

\section{Related Work}

\textbf{Personalized Diffusion Models.} Recent advancements in identity-preserving image generation have seen rapid development, with several approaches employing Textual Inversion \cite{gal2022image}, DreamBooth \cite{ruiz2023dreambooth}, and LoRA \cite{hu2021lora} for fine-tuning models on specific IDs, achieving impressive results. However, these methods lack flexibility and the capability for real-time inference. In contrast, training-free methods effectively eliminate the dependency on parameter fine-tuning. For instance, the IP-Adapter \cite{radford2021learning} leverages CLIP features \cite{radford2021learning} to guide pre-trained models, ensuring identity preservation. PuLID \cite{guo2024pulid} adopts EvaClip \cite{sun2023eva} to maintain identity consistency, while InstantID \cite{wang2024instantid} integrates ArcFace \cite{deng2019arcface} with pose networks to achieve flexible ID retention. Within the realm of identity-preserving video generation, maintaining smooth character motion alongside accurate preservation of identity features represents a key challenge. The ID-Animator \cite{he2024id}, built upon the AnimateDiff \cite{guo2023animatediff} base model, has successfully preserved identity characteristics but exhibits noticeable limitations in the fluidity of character movements. The emerging DiT architecture \cite{yang2024cogvideox, opensora} shows promise for enhancing consistency in video output. ConsisID \cite{yuan2024identity}, which builds on CogVideoX\cite{yang2024cogvideox}, employs a frequency-aware control scheme to enhance identity consistency without the need for identity-specific tuning. However, existing methods still face challenges such as low motion amplitude and facial instability during movement, which can result in limited dynamic movements and unnatural changes or distortions in facial features across frames.

\noindent \textbf{Portrait Animation.} Portrait animation techniques have made significant strides in animating static images by mapping motions and expressions from a guiding video while preserving the portrait's identity and background. \cite{khakhulin2022realistic, wiles2018x2face, yao2020mesh, pang2023dpe, wang2021one} Research primarily focuses on 3D morphable models \cite{blanz2023morphable}, such as DECA \cite{DECA:Siggraph2021} and FLAME \cite{li2017learning}, which excel in detailed 3D face modeling but largely concentrate on facial features without extending to full-body animations or scene elements. In rendering, volumetric methods provide high detail but are computationally intense, while CVTHead \cite{ma2024cvthead} offers a more efficient, yet still facially focused, approach. Animation methods like EchoMimic \cite{chen2024echomimic}, which relies on Mediapipe \cite{lugaresi2019mediapipe}, and SadTalker \cite{zhang2023sadtalker}, which uses audio inputs to generate 3D motion coefficients, also emphasize facial regions. Despite their advancements, these methods generally lack the ability to generate or control complete scenes through text-based inputs, highlighting a gap in creating broader narrative or environmental elements through such interactions.

\section{Methodology}
Given a reference face image, FantasyID is designed to generate a video that faithfully preserves the individual's identity characteristics. An overview of FantasyID is illustrated in Figure \ref{fig:overview}. For each training video, we construct a multi-view face collection and randomly select a reference image as the input condition. Then, we utilize face abstractor to extract 2D clip tokens (Sec. \ref{sec:abstrator}), employ DECA to disentangle features unrelated to the core ID (such as expressions, pose) and to extract 3D structural information (Sec. \ref{sec:blendnet}), and use fusion transformer to fuse the 2D tokens and 3D tokens into face descriptor embeddings (Sec. \ref{sec:fusion_transformer}). Additionally, we exert control over the DiT-based model by employing a layer-aware signal injection method, ensuring precise modulation at each layer (Sec. \ref{sec:injection}). The following section (Sec. \ref{sec:preliminary}) elaborates on the diffusion model and the preliminaries of our method.

\subsection{Preliminary} \label{sec:preliminary}

\textbf{Latent Diffusion Models.} Latent diffusion models are efficient diffusion models that operate in the latent space rather than the pixel space. We use an encoder \( \varepsilon \) from a pre-trained variational autoencoder to compress video data \( x \) into a latent code \( z = \varepsilon(x) \). The encoder \(\varepsilon\) is a video compression module based on 3D variational autoencoders \cite{yu2023language}. It incorporates three-dimensional convolutions to compress videos both spatially and temporally. During the diffusion stage, Gaussian noise $\epsilon$ is added to \( z \) to create \( z_t = \sqrt{\alpha_t} z + \sqrt{1 - \alpha_t} \epsilon \), with \( \epsilon \sim \mathcal{N}(0, I) \), over \( T \) stages. Here, \( \alpha_t \) serves as the noise scheduler, while \( t \) represents the timestep. The denoising process employs the conditional probability \( p_\theta(z_{t-1} | z_t) = \mathcal{N}(\mu_\theta(z_t), \Sigma_\theta(z_t)) \) to predict the previous state \( z_{t-1} \). Here, \(\mu_\theta\) implemented using a denoising model \(\epsilon_\theta\), while \(\Sigma_\theta \) represents the learned covariance of the reverse diffusion process. The training objective typically involves a reconstruction loss that aims to minimize the discrepancy between the added noise and the network's predicted noise: 
{\setlength\abovedisplayskip{7pt}
\setlength\belowdisplayskip{7pt}
\begin{equation}
\mathcal{L} = \mathbb{E}_{t, \mathbf{z} \sim p(z), \epsilon \sim \mathcal{N}(0, I)} \left[\|\epsilon_\theta(\mathbf{z}_t, t) - \epsilon\|_2^2\right]
\end{equation}}

\noindent \textbf{Diffusion Transformer(DiT).} The Diffusion Transformer \cite{peebles2023scalable} is a transformer-based architecture designed for efficient denoising in latent diffusion models. In contrast to to UNet \cite{ronneberger2015u} for processing spatiotemporal latent representations,DiT-based models demonstrate superior capabilities in modeling long-range dependencies and ensuring cross-temporal coherence. This has led to significant advancements in motion coherence and overall video quality \cite{yang2024cogvideox, kong2024hunyuanvideo, opensora}. For our denoising model \(\epsilon_\theta\), we opted for the MM-DiT from CogVideoX \cite{yang2024cogvideox}.

\subsection{Multi-View Collection and Face Abstractor} \label{sec:abstrator}

\textbf{Multi-View Face Collection.} Obtaining effective ID references during the training stage is crucial. To ensure the model focuses on critical areas, we first crop the facial region from each frame of the video, eliminating the background distractions. Following MovieGen \cite{polyak2024movie}, a single reference image can lead the model to learn shortcuts that involve directly replicating the face, so we have constructed a collection of face images \( \mathcal{I} \) from different viewpoints in the  training stage. We utilize RetinaFace \cite{deng2020retinaface} to extract geometric relationships among facial landmarks to calculate the head pose angles and select six images with the most significant viewpoint differences to form a multi-view face set. By providing the model with a diverse range of perspectives, it gains a more comprehensive understanding of the subject, thereby enhancing its ability to maintain identity consistency across various poses and expressions.

\noindent \textbf{Face Abstractor.} After acquiring the face image from the multi-view face collection, we aim to effectively extract facial features from it. Previous works \cite{yuan2024identity, he2024uniportrait} on identity-preserving generation has employed Q-Former \cite{li2023blip} to transform the face clip embeddings, but this approach can disrupt the spatial structure among features \cite{cha2024honeybee}. Recognizing the critical importance of local correlations for face characteristics, we introduce the use of C-Abstractor \cite{cha2024honeybee} to transform face clip embeddings to \( X_f \in \mathbb{R}^{h \times w \times C} \). This module, composed of two-dimensional convolutions and average pooling, leverages spatial locality to enhance feature extraction. By effectively capturing comprehensive facial information while preserving the key spatial relationships essential for accurate video generation.

\subsection{3D Constraints} \label{sec:blendnet}

\textbf{Face Vertex.} A stable facial geometric constraint is critical to ensuring high-quality generation. We utilize the priors obtained from 3D reconstruction to both constrain and enhance the model's understanding of the reference image, thereby improving the quality and fidelity of the generated output. We employ the DECA framework \cite{feng2021learning} to capture the three-dimensional geometric structure of faces, which provides vertex coordinates \( V_s  \in \mathbb{R}^{(N \times 3)} \) for the reference image, where \(N\) is the number of vertices in FLAME model \cite{li2017learning}. This approach distinguishes between intrinsic facial features and extrinsic factors such as pose, expression, and lighting. This separation mechanism enhances the model's comprehension of identity-specific features while effectively suppressing interference from non-identity characteristics.

\noindent \textbf{3D Vertex Representation.} We employ SpiralNet++ \cite{gong2019spiralnet++} to extract 3D features from \( V_s \), represented as \( X_V' \in \mathbb{R}^{N' \times C'} \), where \( N' \) denotes the number of vertices after downsampling and \( C' \) indicates the channel dimension of the feature descriptor. To encode positional information from the 3D point cloud, we incorporate a positional encoding \( E_{\text{dep}} \), derived from the depth map of the projected vertices. This results in the enhanced vertex features is \( X_V = X_V' + E_{\text{dep}} \).

\subsection{Fusion Transformer} \label{sec:fusion_transformer}

To effectively integrate 3D point cloud features with complementary 2D descriptors, we design the fusion transformer. Specifically, We utilize MLP to transform the 2D features \( X_f \in \mathbb{R}^{h \times w \times C} \) into \( X_F \in \mathbb{R}^{h \times w \times C'} \), aligning the feature dimension from \( C \) to \( C' \). The aligned 2D features are then concatenated with the 3D vertex features, formulated as \( X = [X_V, X_F] \), where the combined feature \(X \in \mathbb{R}^{(N' + h \times w) \times C')} \). Here, \( h \times w \) denotes the dimensions of the 2D feature space. The fusion transformer consists of \(L\) layers. Finally, we utilize a series of residual 1D convolutions, referred to as ResConv1D, to align the hidden dimensions of DiT, thereby obtaining the id descriptor \( V_F \in  \mathbb R^{(N' \times C)}\). Through this fusion, we effectively integrates high-dimensional data into a integrated features, capturing rich facial representations while ensuring the preservation of 3D structural priors in the generated facial representations.

\subsection{Layer-Aware Control Signal Injection} \label{sec:injection}

Inspired by the observation that each layer in the DiT architecture contributes uniquely to the overall performance \cite{yu2024representation}, we adopt a similar approach for controlling facial video generation using DiT. Specifically, we recognize that different layers exhibit varying sensitivities to control signals. To address this, we propose a layer-aware control signal injection mechanism that dynamically adjusts the integration of control signals based on the role of each layer.

Particularly, for each MM-DiT block, we employ a lightweight model 
${F}_l$ to learn the optimal feature representation. This lightweight network comprises a convolution block followed by normalization. Independent weights for each layer enhance fidelity and diversity, aligning control signals precisely with the needs of each layer. This ensures stability and expressive potential in outputs. The process is defined by the formula:
{\setlength\abovedisplayskip{7pt}
\setlength\belowdisplayskip{7pt}
\begin{equation}
\hat{Z_l} = Z_l + {F}_l(\text{Attention}(l, Q_l, K_l^{\text{id}}, V_l^{\text{id}}))
\end{equation}}

where \(Q_l, K_l^{id}, V_l^{id} \) are the query, key, and values matrices of the attention operation, \( Q_l = Z_l W_l^q \), \( K_l^{\text{id}} = V_F W_l^k \), \( V_l^{\text{id}} = V_F W_l^v \). Here, \( Z_l \) represents the hidden states, and \( l \) is the layer index of the MM-DiT block, and \( W_l^q, W_l^k, W_l^v \) are trainable weights.

\begin{figure*}[h!]
    \centering
    \includegraphics[width=18cm]{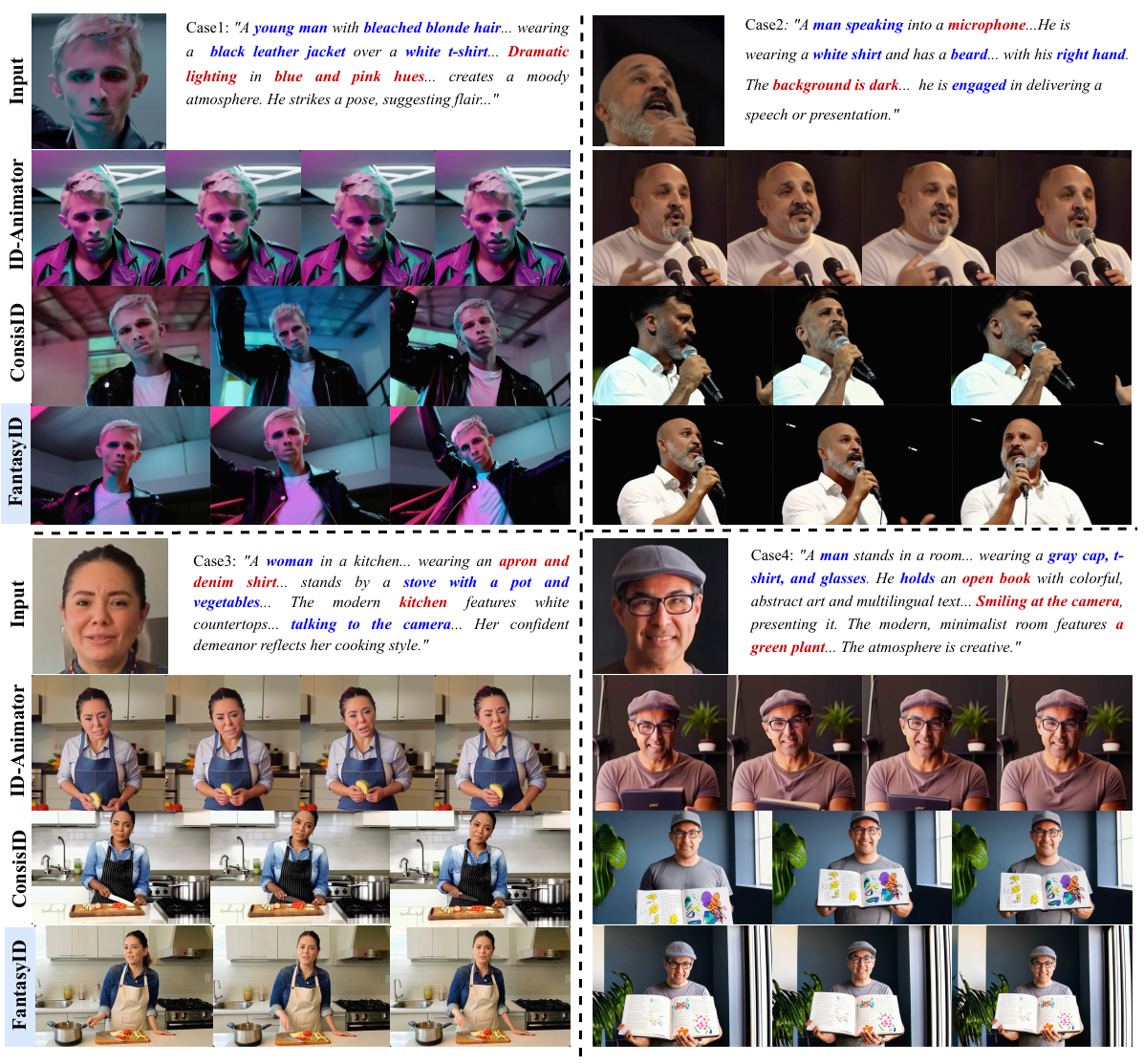}
    \caption{Qualitiative Comparision between our methods and ConsisID, ID-Animator. Please refer our supplementary materials for the video results.}
    \label{fig:qualitative}
\end{figure*}
\section{Experiments}
\subsection{Setups}

\textbf{Implementation Details.} In our experiments, we utilize a diverse dataset  comprising full body and portrait data from ConsisID-Data \cite{yuan2024identity}, CelebV-HQ\cite{zhu2022celebvhq}, and Open-Vid\cite{nan2024openvid}. Subsequently, following the approach in SVD \cite{blattmann2023stable}, we employed PaddleOCR \cite{zhou2017east,liao2022real} to eliminate any videos containing subtitles. Furthermore, we used InsightFace \cite{deng2019arcface,Deng2020CVPR} to exclude videos with a face confidence score below 0.9, resulting in a final selection of approximately 50,000 clips. We optimize with a batch size of 16 and a learning rate of \(3 \times 10^{-6}\), completing 90,000 training steps, which takes approximately 36 hours using 16 A100 GPUs. Our methodology incorporates classification-free guidance with a random null text ratio of 0.1, utilizing AdamW as the optimizer, and employs a cosine with restarts as the learning rate scheduler. During the inference stage, we utilize DECA's coarse FLAME \cite{li2017learning} parameters to construct a 3D point cloud from the input image. The denoising step is set to 50. The fusion transformer is designed to 6 layers, and the ResConv1D comprises 4 residual 1D convolutional blocks. The downsampling factor of the face abstractor is set to 4, and the number of 3D tokens \( N' \) is 314.

\noindent \textbf{Evaluation Metrics.} We employ ArcFace \cite{deng2019arcface} embedding similarity to assess two key aspects. First, Reference Similarity(RS) calculates the similarity between the reference image and frames to evaluate identity preservation. Second, Inter-Frame Similarity(IFS) measures the similarity between consecutive video frames to evaluate the stability of identity features during motion. Additionally, we analyze the Fréchet Inception Distance (FID) \cite{heusel2017gans} of the face region to assess video quality and utilize Face Motion (FM), measured by average dense optical flow \cite{farneback2003two}, to evaluate the degree of motion. We used 50 richly detailed portrait reference images. To more accurately measure the identity preservation capability, we cropped the facial regions from each video for quantitative evaluation.

\subsection{Qualitative Analysis}

For the qualitative evaluation, we present comparison results with diffusion-based identity preservation models, ConsisID and ID-Animator. Other models, including VideoMaker \cite{wu2024videomaker} and MagicMirror \cite{comas2024magicmirror}, are not open-source and therefore not included in our direct comparisons. 

Figure \ref{fig:qualitative} demonstrates that ID-Animator struggles with generating human body parts beyond the face and exhibits noticeable ``copy-paste" artifacts. Moreover, the generated content often appears overly static, lacking natural motion. These limitations significantly restrict its practical application in scenarios requiring dynamic and realistic human behavior or interactions. Regarding ConsisID, while the overall visual quality is high, there are still issues with structural instability during facial movements, as seen in Case 1. Although ConsisID retains features such as skin texture from the reference image, it fails to accurately reproduce the overall facial structure in Case 3 and 4. In contrast, our method achieves the best results in terms of visual quality, preservation of the subject's identity from the reference image, and maintaining consistent facial structure across frames during motion. 

To further validate the effectiveness of our proposed method, we conducted a comprehensive user study involving 32 participants. Each participant was tasked with assessing four critical aspects: Overall Quality(OQ), Face Similarity(F-Sim), Facial Structure(F-Str), and Facial Dynamics(FD), rating each aspect on a scale from 0 to 10. As shown in Table \ref{tab:userstudy}, the scores indicate that FantasyID consistently outperforms baseline methods across all evaluated dimensions, demonstrating its superior perceived quality in human assessments.

\begin{table}[ht]
    \centering
    \small
    \begin{tabular}{>{\raggedright\arraybackslash}p{1.7cm}*{4}{>{\centering\arraybackslash}p{1.08cm}}}
        \toprule
        & \textbf{OQ} & \textbf{F-Sim} & \textbf{F-Str} & \textbf{FD} \\
        \midrule 
        ID-Animator & 4.38 &  6.20 & 5.82 & 3.28 \\
        ConsisID    & 7.85 & 7.79 & 6.44  &  7.12\\
        Ours        & \textbf{8.39} &  \textbf{8.68}  & \textbf{8.10} & \textbf{7.94}   \\
        \bottomrule
    \end{tabular}
    \caption{\textbf{User Study results.} Higher scores indicate better performance.}
    \label{tab:userstudy}
\end{table}

\subsection{Quantitative Analysis}

\begin{figure}[h!]
    \centering
    \includegraphics[width=8.5cm]{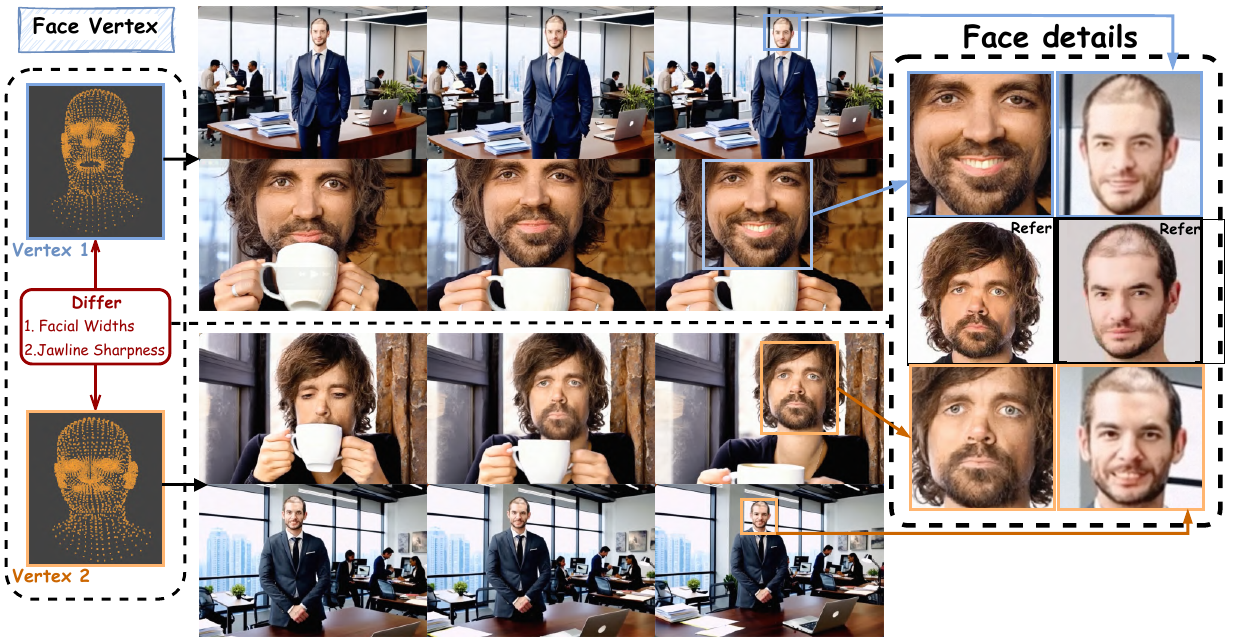}
    \caption{\textbf{Effect of 3D Constraints.} By altering the facial widths and jawline sharpness of face vertex, the generated facial videos exhibit noticeable structural changes.}
    \label{fig:effect3d}
\end{figure}

Table \ref{tab:quantivate} presents a comprehensive quantitative evaluation of various face video generation methods. ID-Animator achieves an impressive FID score and a higher IFS score. However, this performance can be attributed to its tendency to generate more static content, thereby ensuring high quality and excellent identity consistency. This focus on static representations likely limits its ability to produce diverse and dynamic facial motions. In contrast, while our method provides a slightly highest FID score, it excels in capturing dynamic expressions, as evidenced by the leading face motion score of 0.61, and achieves the highest RS score of 0.57, reflecting superior identity preservation. Notably, our model outperforms ConsisID across all metrics, reflecting a superior ability in dynamism and identity preservation.

\begin{table}[ht]
    \centering
    \small
    \begin{tabular}{>{\raggedright\arraybackslash}p{1.7cm}*{4}{>{\centering\arraybackslash}p{1.08cm}}}
        \toprule
        & \textbf{FID }$\downarrow$ & \textbf{RS}$\uparrow$ & \textbf{IFS}$\uparrow$ & \textbf{FM}$\uparrow$  \\
        \midrule 
        ID-Animator & \textbf{138.27} & 0.35 & \textbf{0.98} &  0.18 \\
        ConsisID    & 149.70 & 0.47 & 0.93 &  0.54 \\
        Ours        & 142.50 &  \textbf{0.57}    &   0.95    &    \textbf{0.61}  \\
        \bottomrule
    \end{tabular}
    \caption{\textbf{Quantitative evaluation of different methods.} The best results are highlighted
in bold.}
    \label{tab:quantivate}
\end{table}

\subsection{Ablation Studies}
To comprehensively evaluate the contribution of each module within the FantasyID framework, we conducted a series of ablation studies. These experiments systematically removed individual components to assess their impact on the overall performance of the model in Table \ref{tab:tab2}. Specifically, we examined the effects of excluding the Multi-View Face Collection(MFC), Face Abstractor(FA), Face Vertex(FV), and Layer-Aware Control Signal Injection (LACSI). Additionally, we modify the face vertex data of different inputs to validate the effectiveness of 3D constraints.

\begin{table}[ht]
    \centering
    \small
    \begin{tabular}{>{\raggedright\arraybackslash}p{2.1cm}*{4}{>{\centering\arraybackslash}p{1.1cm}}}
        \toprule
        & \textbf{FID } $\downarrow$ & \textbf{RS} $\uparrow$ & \textbf{IFS} $\uparrow$ & \textbf{FM} $\uparrow$  \\
        \midrule 
        w/o FA  & 145.82 & 0.55 & 0.93 & 0.49 \\ 
        w/o MFC   & \textbf{130.77} & 0.54 & \textbf{0.98} &  0.33 \\
        w/o FV & 172.51 & 0.42 & 0.90 &   0.36 \\
        w/o LACSI & 235.46 & 0.33 & 0.93 & 0.22 \\
        FantasyID & 142.50 &  \textbf{0.57} & 0.95 & \textbf{0.61}  \\
        \bottomrule
    \end{tabular}
    \caption{\textbf{Quantitative results of removing individual modules from FantasyID framework.}}
    \label{tab:tab2}
\end{table}

\begin{figure*}[h!]
    \centering
    \includegraphics[width=17.0cm]{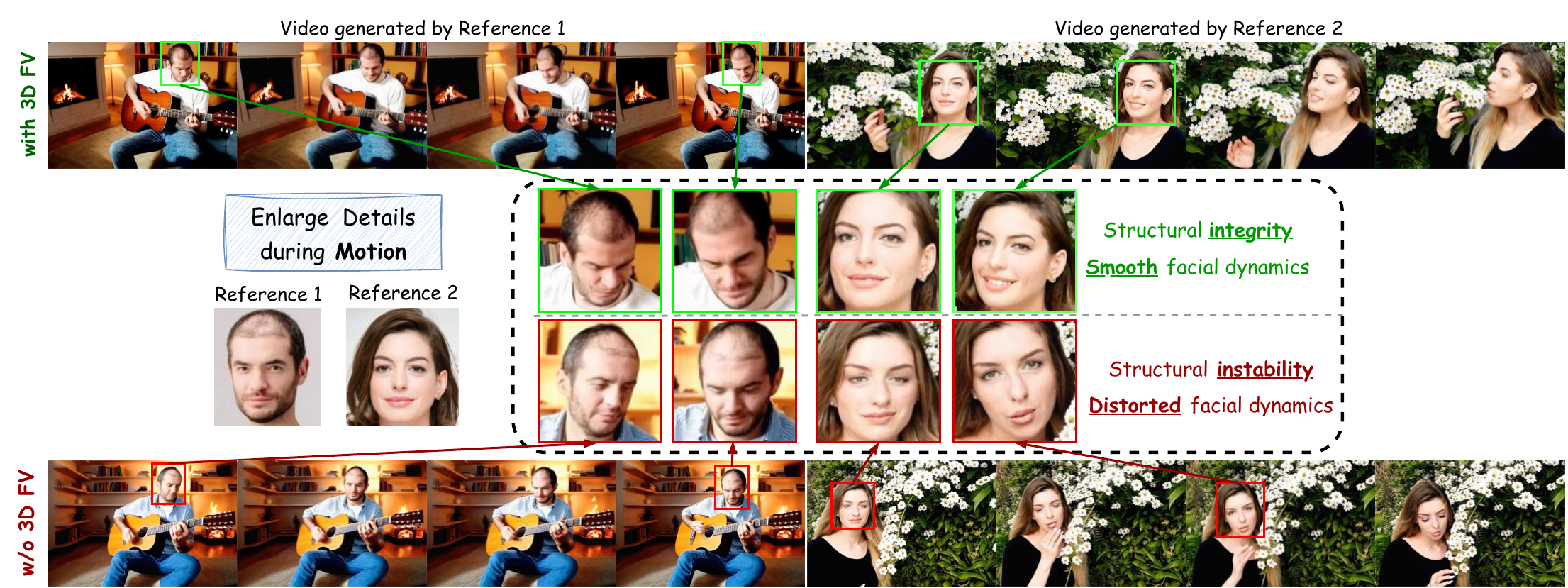}
     \captionsetup{width=18cm}
    \captionof{figure}{\textbf{Ablation study on Face Vertex(FV)}. Without the face vertex leading to distorted facial structures during motion.}
    \label{fig:wo_3d}
\end{figure*}

\begin{figure}[h!]
    \centering
    \includegraphics[width=8.5cm]{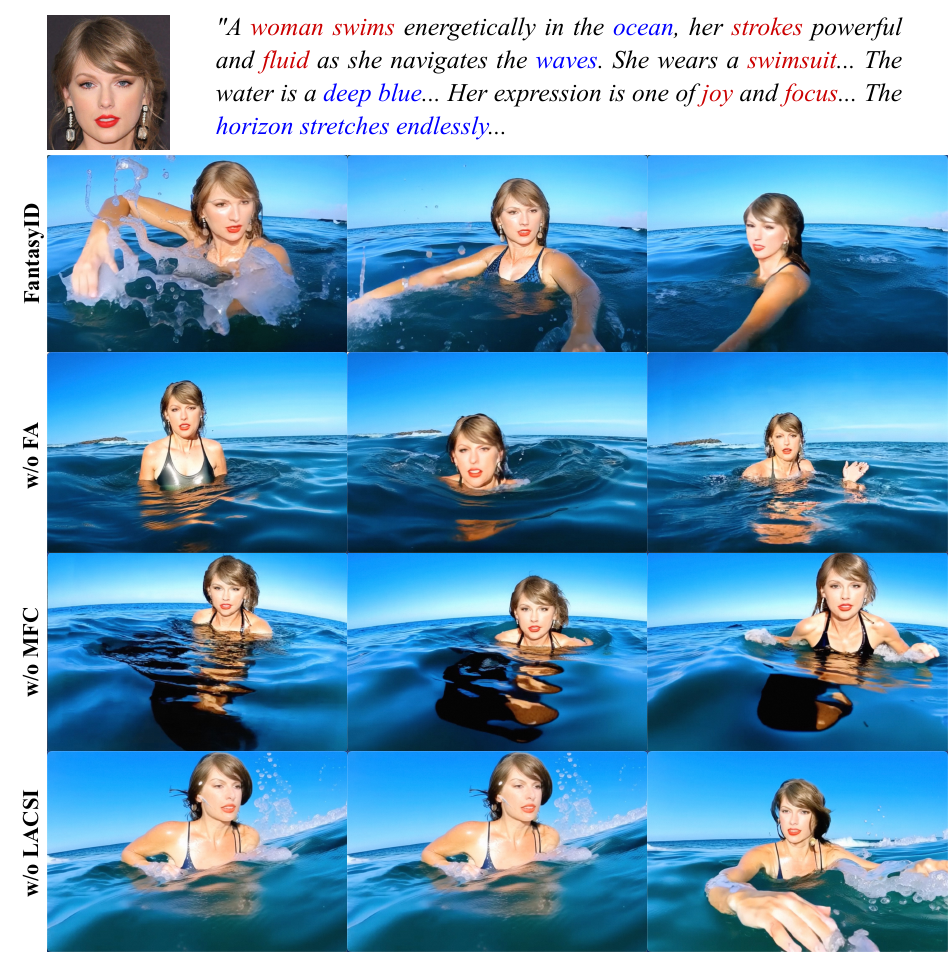}
    \caption{\textbf{Ablation studies on Multi-View Face Collction(MFC), Face Abstractor(FA) and Layer-Aware Control Signal Injection(LACSI).}}
    \label{fig:ab_multi}
\end{figure}
\vspace{-0.1cm}

\subsection{Qualitative Analysis}
\noindent \textbf{Effect of 3D Constraints.} To verify the efficacy of our 3D constraint control mechanism, we modify the 3D face vertex to generate videos with different facial widths and jawline sharpness. The qualitative results presented in Figure \ref{fig:effect3d} showcase the significant variations in facial structure, thereby confirming that our 3D constraints effectively guide the generation of facial features. This demonstrates the flexibility and precision of our approach in controlling facial characteristics.

\noindent \textbf{w/o Face Vertex.} We evaluated the importance of 3D constraints by excluding the face vertex of our framework. The qualitative results in Figure \ref{fig:wo_3d} demonstrates that the absence of 3D face vertex data causes the model to rely solely on 2D feature extraction, leading to distortions in facial structure during motion. The quantitative results in Table \ref{tab:quantivate} show a decline across all metrics, which suggest more erratic facial motion.  These results indicate that the critical role of 3D vertex integration in preserving structural integrity and ensuring smooth facial dynamics.

\noindent \textbf{w/o Multi-View Face Collection.} We replace the multi-view face collection with a single face image during the training stage. As shown in Figure \ref{fig:ab_multi}, this approach significantly reduces the range of captured facial motions, limiting the model's ability to understand and represent different angles. However, this approach achieves the best FID and IFS scores as shown in Table \ref{tab:tab2}. This performance can be attributed to the model's tendency to take a shortcut by prioritizing higher similarity to the reference image, thereby maintaining consistency at the expense of dynamic range.

\noindent \textbf{w/o Face Abstractor.} We replaced the face abstractor with Q-Former, which, as shown in Figure \ref{fig:ab_multi}, leads to some facial distortions. These distortions are likely due to Q-Former's tendency to disrupt the spatial characteristics of face CLIP features. Additionally, the results presented in Table 2 indicate that this approach achieves lower FID, RS, and IFS scores. This suggests that  Face Abstractor is more effective at capturing comprehensive and spatially coherent facial information.

\noindent \textbf{w/o Layer-Aware Control Signal Injection.} By removing the layer-aware control signal injection module \(F_l\), we observed a significant decrease in face similarity, as shown in Figure \ref{fig:ab_multi}, along with a decline in all metric scores, as detailed in Table \ref{tab:tab2}. These results indicate a decline in both video quality and identity preserving. In contrast, the layer-aware control method adapts more effectively to the unique feature distributions between different DiT blocks by learning the most suitable feature control signals for each layer. This approach ensures optimal performance and fidelity in generating ID features.

\section{Conclusion}

FantasyID presents a groundbreaking approach for identity-preserving human video generation, overcoming the limitations of traditional methods. By employing a multi-view face collection, face abstractor, 3D constraints, and layer-aware control signal injection, it significantly enhances video quality, identity preserving, and temporal coherence. This scalable, training-free solution maintains high-fidelity representations during complex motions. Future work will focus on optimizing multi-identity retention and expanding FantasyID's role in dynamic video production and personalized content creation.

\clearpage
\printbibliography
\end{document}